\documentclass[pre,aps,showpacs,twocolumn]{revtex4-1}
\usepackage{graphicx}
\usepackage{pstricks}
\usepackage{amssymb}
\usepackage{amsmath}
\usepackage{amsfonts}
\usepackage{natbib}
\usepackage{bm} 

\graphicspath{{figure/}}
\newcommand{\beq}{\begin{equation}}
\newcommand{\eeq}{\end{equation}}
\newcommand{\bea}{\begin{eqnarray}}
\newcommand{\eea}{\end{eqnarray}}
\newcommand{\ra}{\rightarrow}

\renewcommand{\r}{\right \rangle}
\renewcommand{\l}{\left \langle}
\newcommand{\nt}[1]{\mathrm{#1}}
\newcommand{\ntbm}[1]{\bm{\mathrm{#1}}}

\makeatletter
\def\blfootnote{\gdef\@thefnmark{}\@footnotetext}
\makeatother

\begin{document}

\title{The unlikely Carnot efficiency}
\date{\today}
\author{Gatien Verley$^1$}
\author{Massimiliano Esposito}
\affiliation{Complex Systems and Statistical Mechanics, University of Luxembourg, L-1511 Luxembourg, G.D. Luxembourg}
\author{Tim Willaert} \author{Christian Van den Broeck}
\affiliation{Hasselt University, B-3590 Diepenbeek, Belgium}

\maketitle

\textbf{The efficiency of an heat engine is traditionally defined as the ratio of its average output work over its average input heat. Its highest possible value was discovered by Carnot in 1824 and is a cornerstone concept in thermodynamics. It led to the discovery of the second law and to the definition of the Kelvin temperature scale. 
Small-scale engines operate in presence of highly fluctuating input and output energy fluxes. They are therefore much better characterized by fluctuating efficiencies. In this letter, using the fluctuation theorem, we identify universal features of efficiency fluctuations. While the standard thermodynamic efficiency is, as expected, the most likely value, we find that the Carnot efficiency is, surprisingly, the least likely in the long time limit. Furthermore the probability distribution for the  efficiency assumes a universal scaling form when operating close-to-equilibrium. We illustrate our results analytically and numerically on two model systems.}

\section{Introduction}

The efficiency of an heat engine operating between a hot and cold reservoir is defined by $\eta=-W/Q_\nt{h}$, where $Q_\nt{h}$ is the heat extracted from the hot reservoir, $-W$ is the work produced by the engine and $-Q_\nt{c}$ is the remaining heat dumped into the cold reservoir. Throughout the letter energy fluxes such as work or heat are positive when flowing into the engine. The Carnot efficiency $\eta_\nt{C}=1-T_\nt{c}/T_\nt{h}$ is easily found to be the upper bound of the engine efficiency, $\eta \leq \eta_\nt{C}$, by combining the first and second law of thermodynamics:
\begin{eqnarray}
\Delta E &=& W+Q_\nt{h}+Q_\nt{c},  \\ 
\Delta S_{\nt{tot}}&=& \Delta S -Q_\nt{h}/T_\nt{h}-Q_\nt{c}/T_\nt{c} \geq 0, 
\end{eqnarray}
in which the engine energy change $\Delta E$ and entropy change $\Delta S$ have to be set equal to zero, as the engine returns to its original state after each cycle.

Early on, Maxwell raised questions concerning the validity of the second law at small scales \cite{Maxwell1878_vol17}. Szilard discussed similar issues in the context of information processing \cite{Szilard1929_vol53}. 
The theoretical breakthroughs in fluctuation theorems and stochastic thermodynamics have fully clarified these points and enable a consistent thermodynamic description of small-scale systems operating arbitrarily far-from-equilibrium \cite{Jarzynski2011_vol2, Sevick2008_vol59, Campisi2011_vol83, Seifert2012_vol75, VandenBroeck2014_vol, Esposito2009_vol81, Liphardt2002_vol296, Blickle2006_vol96, Kung2013_vol113}. 
These developments are of crucial practical relevance nowadays, as we are able to design machines operating at the nanoscale as well as to study in great detail biological machines transducing energy and processing information at the sub-micron scale \cite{Saira2012_vol109, Blickle2012_vol8, Moffitt2009_vol457, Yasuda2001_vol410, Toyabe2010_vol6, Berut2012_vol483, Alemany2012_vol8, Collin2005_vol437, Koski2013_vol9, Kung2012_vol2, Ciliberto2013_vol110, Bustamante2005_vol58, Matthews2013_vol}. For such engines, fluctuations are ubiquitous and quantities such as work $w$, heat $q$, energy change $\Delta e$, entropy change $\Delta s$ and entropy production $\Delta s_\nt{tot}$ are stochastic and contain a much richer information than their ensemble average values work $W=\langle w \rangle$, heat $Q=\langle q \rangle$, energy change $\Delta E= \l \Delta e \r$, etc.
At the stochastic level, the first law is essentially the same as at the average level, while the second law of thermodynamics is replaced by a universal symmetry in the probability distribution for the total entropy called fluctuation theorem \cite{Jarzynski2011_vol2, Sevick2008_vol59, Campisi2011_vol83, Seifert2012_vol75, VandenBroeck2014_vol, Esposito2009_vol81}:
\begin{eqnarray}
&& \Delta e = w+q_h+q_\nt{c}, \;\; \Delta s_\nt{tot}=\Delta s -q_\nt{h}/T_\nt{h}-q_\nt{c}/T_\nt{c} \label{eq:stochlaws},\\
&&\hspace{1.5cm} \frac{P(\Delta s_\nt{tot})}{P(-\Delta s_\nt{tot})}=\exp(\Delta s_\nt{tot}),\label{s3}
\end{eqnarray}
where the Boltzmann constant is set to unity ($k_B = 1$). In words, the probability for observing a trajectory with entropy increase $\Delta s_\nt{tot}$ is exponentially more likely than the probability to observe the corresponding entropy decrease. The second law follows by Jensen's inequality, more precisely by taking the logarithm of the following inequality $1 = \l e^{-\Delta s_\nt{tot}} \r \ge e^{- \l \Delta s_\nt{tot} \r }$. The fluctuation theorem (\ref{s3}) is a probabilistic statement connecting the energy fluxes in the engine since the entropy production can be expressed with them. It is surprising that, in view of this dramatic reformulation of the second law, and despite its founding role in thermodynamics, the properties of the resulting stochastic efficiency $\eta \equiv - w/q_\nt{h}$ have not yet been explored. We will do so in this letter and identify universal features of the corresponding probability distribution $P_t(\eta)$. Note that at equilibrium all realizations are reversible, i.e. $\Delta s_\nt{tot}=0$, leading to a stochastic efficiency equal to the Carnot efficiency $\eta=\eta_\nt{C}$. When operating irreversibly, it follows from the fluctuation theorem (\ref{s3}) that realizations with both $\Delta s_\nt{tot}> 0$ and $\Delta s_\nt{tot}< 0$ appear, the latter albeit with a probability which is exponentially smaller. Hence efficiencies lower but also higher than Carnot will be observed. 


\section{Results}

\subsection{Properties of efficiency fluctuations}

Our most striking result is the following: the Carnot efficiency $\eta_\nt{C}$ becomes the least likely efficiency for long times as a direct consequence of the fluctuation theorem (\ref{s3}).
This result holds for engines with finite state space and therefore with bounded energy so that the energy and entropy contributions $\Delta e$ and $\Delta s$  can be neglected in the first and second law in the long time limit as compared to the work and heat. Hence, the heat dumped in the cold reservoir is just $-q_\nt{c}=w+q_\nt{h}$ and we can focus on the work and heat variables, $w$ and $q=q_\nt{h}$, and their corresponding output and input powers, $-\dot{w}\equiv-w/t$ and $\dot{q} \equiv q/t$.
For systems that harbour no long-time correlations, work $\dot{w}$, heat $\dot{q}$ and their ratio $\eta=-{w}/{q}=-\dot{w}/\dot{q}$ are expected to converge in the infinite time limit to their most probable value values $\l w\r/t$, $\l q \r /t$ and $\bar \eta \equiv -\l w \r /\l q\r$. The latter efficiency is the one predicted by standard thermodynamics, with the Carnot efficiency as upper bound: $\bar \eta\leq \eta_\nt{C}$. When considering long trajectories, the probability distribution at time $t$ for $w$ and $q$ as well as for $\eta$ are described by the theory of large deviations \cite{Touchette2009_vol478} and assume the following asymptotic form:
\beq
P_t(\dot{w}, \dot{q}) \sim  e^{-t I( \dot{w}, \dot{q})},\;\;\; P_t(\eta) \sim e^{-t J(\eta)}.
\label{eq:Proba}
\eeq
The so-called large deviation functions $I(\dot{w}, \dot{q})$ and $J(\eta)$ describe the exponentially unlikely deviations of $\dot{w}$, $\dot{q}$ and $\eta$ from their most probable values. These functions vanish at the most likely value of the probability distribution. They are otherwise strictly positive. In particular, we have $J(\bar \eta)=0$.

The large deviation function $J(\eta)$, being the ratio of $-\dot{w}$ over $\dot{q}$, is obtained from $I$ by the following so-called contraction: 
\beq
J(\eta) = \min_{\dot{q}} I(-\eta \dot{q}, \dot{q}). \label{eq:defJeta}
\eeq 
Intuitively, the decay rate $J$ for a given efficiency is the smallest among all the decays rates $I$ for the input and output powers which reproduce this efficiency. Note that since the minimization over $\dot{q}$ in (\ref{eq:defJeta}) includes $\dot{q}=0$, we find that $J(\eta) \leq I(0,0)$.
   
Expressing the entropy production as $\Delta s_\nt{tot}=q(1/T_\nt{c}-1/T_\nt{h}) +w/T_\nt{c}$, the fluctuation theorem (\ref{s3}) can be shown to assume the more detailed form \cite{Garcia-Garcia2010_vol82,Sinitsyn2011_vol44}: 
\begin{eqnarray}
\frac{P(w,q)}{P(-w,-q)}=\exp(\Delta s_\nt{tot}).\label{wq}
\end{eqnarray}
Reversible realizations are characterized by $\Delta s_\nt{tot}=0$ and thus by an efficiency equal to Carnot efficiency $\eta=-\dot{w}/\dot{q}=\eta_\nt{C}$. In this case, $P(w,q)=P(-w,-q)$ implying $I(\dot{w},\dot{q})=I(-\dot{w},-\dot{q})$. Hence $I(-\eta_\nt{C} \dot{q}, \dot{q})$ is a symmetric function of the input power $\dot{q}$. Since $I$ is generically a convex function (assuming no phase transitions),  the minimum in equation (\ref{eq:defJeta}) is, for $\eta=\eta_\nt{C}$, reached in $\dot{q}=0$ and thus $J(\eta_\nt{c}) = I(0,0)$. Since $J(\eta) \leq I(0,0)$, this proves our main result, namely that the Carnot efficiency becomes the least likely when $t\rightarrow \infty$: $J(\eta) \le J(\eta_\nt{C})$ or $P_t(\eta) \ge P_t(\eta_\nt{C})$.

Summarizing so far, all values of the efficiency are possible in a small scale engine running for a large but finite time, including those forbidden by the second law at the average level. The probability for an efficiency different from the standard thermodynamic value $\bar \eta$ decreases exponentially with time with the strongest decrease observed for the Carnot efficiency. Therefore, the probability distribution for the efficiency will develop an exponentially pronounced minimum at the Carnot efficiency as one monitors longer operation times. This observation provides a novel way to define the temperature scale. In standard thermodynamics, the Kelvin temperature scale is introduced by the measurement of the Carnot efficiency of a reversible engine, measurement which is in principle unattainable in a finite time.  The identification of the least likely efficiency in a system operating away from reversibility provides an alternative measurement of the  Kelvin temperature, which does not suffer from this predicament.	
For a machine designed to operate like an heat engine (``on average''), the rare events leading to stochastic efficiencies larger than the Carnot efficiency correspond to realizations along which the machine functions as a heat pump, while those with efficiency lower than zero correspond to a dud engine dissipating heat while absorbing work. 
The above derivation has been illustrated on an heat engine operating in continuous time,  but the results remain valid for all other types of machines, such as isothermal energy transducers, heat engines, refrigerators or heat pumps, operating in non-equilibrium steady-states or cyclically as long as the driving cycle is invariant under time-reversal. The only difference is that the Carnot efficiency has to be replaced by the reversible efficiency. Below, we give an example of an isothermal work-to-work conversion where the reversible efficiency is equal to $1$ and thus corresponds to the least likely efficiency.

Beyond these striking general conclusions about the least and most likely efficiency, we proceed to show that the efficiency fluctuations in the close-to-equilibrium regime have a universal scaling form. Our starting point is that close-to-equilibrium, the relevant work $w$ and heat $q$ fluctuations are generically Gaussian. 
The resulting large deviation function for the efficiency, being the ratio of two correlated Gaussian variables, is found to be:
\beq
J(\eta) =\frac{1}{2} \frac{ \left (\eta \l \dot q \r + \l \dot w\r\right )^2}{\eta^2 C_{qq}  + 2  \eta C_{wq} +  C_{w w}}, \label{eq:CloseEqLDFeta}
\eeq
where $C_{wq} \equiv (\l wq \r - \l w\r\l q\r)/t$, $C_{ww}$ and $C_{qq}$ are the elements of the symmetric covariance matrix. 
The crucial thermodynamic ingredient is obtained by combining the Gaussian statistics for $\dot w$ and $\dot q$  with the fluctuation theorem. More precisely, noting that  (\ref{wq}) has to be valid for all values of $w=t\dot w$ and $q=t\dot q$, one finds (cf. methods section): 
\beq
\l \dot q \r  = \frac{\eta_\nt{C} C_{qq} + C_{wq} }{2T_\nt{c}}, \;\;\;
\l \dot w \r = \frac{\eta_\nt{C} C_{wq} +C_{ww} }{2T_\nt{c}}. \label{eq:ac}
\eeq
The large deviation function $J$ can therefore be solely expressed in terms of the covariance matrix:
\beq
J(\eta) =\frac{1}{8 T_\nt{c}^2} \frac{[\eta \eta_\nt{C} C_{qq} + (\eta+\eta_\nt{C})C_{wq}+C_{ww}]^2}{\eta^2 C_{qq} + 2\eta C_{wq}+C_{ww}}.
\label{eq:JetaCloseEq}
\eeq 
The above explicit expression for $J(\eta)$ is in agreement with the general properties pointed out above, namely its minimum and maximum are reached for $\eta =\bar \eta$ and $\eta=\eta_\nt{C}$, respectively. These are also the two only extrema of the function.  
Remarkably, the least likely decay rate, i.e. the rate at Carnot efficiency, can be rewritten from (\ref{eq:ac},\ref{eq:JetaCloseEq}) solely in terms of the average heat and work: $J(\eta_\nt{C}) =  (\eta_\nt{C}   \l \dot q \r + \l \dot w \r)/(4T_\nt{c})$. This relation, which ultimately derives from the fluctuation theorem, should be easy to test experimentally. 
We also note that the two asymptotic values of $J(\eta)$ at $\eta \rightarrow \pm \infty$ coincide, namely $J( \infty)= (\eta_\nt{C} C_{qq}+ C_{wq})^2 /(8T_\nt{c}^2 C_{qq})=  \l \dot q \r^2/(2C_{qq})$. We used (\ref{eq:ac}) for the second equality. 
Since the covariance matrix is directly related to the Onsager matrix (cf. methods section), this latter can be obtained from measurements of efficiency fluctuations close to equilibrium. 
In fact the covariance matrix $C_{ww}$, $C_{wq}$ and $C_{qq}$ is uniquely specified by the most probable efficiency $\bar \eta$, the value of LDF at the Carnot efficiency $J(\eta_\nt{C})$, and the asymptotic value of the LDF $J(\infty)$ (cf. method section), so that (\ref{eq:JetaCloseEq}) can be rewritten as:
\beq\label{eq:scaling}
\frac{J(\eta)}{J(\eta_\nt{C})} = \frac{(\bar \eta-\eta)^2}{ (\bar \eta -2\eta+\eta_\nt{C})(\bar \eta - \eta_\nt{C}) + \frac{J(\eta_\nt{C})}{J(\infty)}(\eta-\eta_\nt{C})^2}  .
\eeq

\subsection{Brownian work-to-work converter}
 
\begin{figure}
\includegraphics[width=6cm]{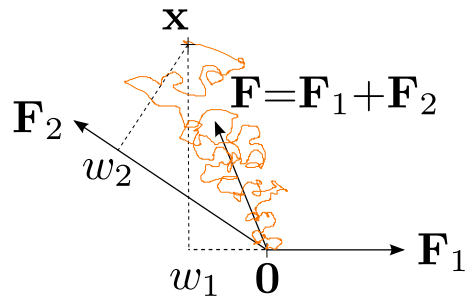}
\caption{Work-to-work converter for an overdamped Brownian particle diffusing in a plane. The particle is driven by the two forces $\ntbm F_1$ and $\ntbm F_2$. The orange line is a specific trajectory ending at position $\ntbm x$. The stochastic work $w_1$ and $w_2$ are obtained from the scalar products (dashed lines) projecting $\ntbm x$ on $\ntbm F_1$ and $\ntbm F_2$ respectively. The ratio of the lengths obtained from these projections gives the stochastic efficiency for this specific trajectory. \label{fig1}}
\end{figure}
As a first illustration of our main results, we consider the simplest possible model for work-to-work conversion \cite{Astumian2009_vol113}. An overdamped Brownian particle subjected to two constant forces $\ntbm F_1$ and $\ntbm F_2$ diffuses on a plane, as illustrated in Fig.~\ref{fig1}. $\ntbm F_2$ is the driving force, allowing the particle to move against an opposing force $\ntbm F_1$. 
For a given displacement $\ntbm x= \ntbm x (t)$ of the particle (assuming $\ntbm x (0) = \ntbm 0$), the work performed by each force is given by $w_1 = \ntbm F_1 \cdot \ntbm x$ and $w_2 = \ntbm F_2 \cdot \ntbm x$. The corresponding stochastic efficiency is $\eta = -w_1/w_2$. 
The displacement $\ntbm x (t)$ is a Gaussian random variable with average $\l \ntbm x (t) \r = \mu \ntbm F t$, where $\ntbm F=\ntbm F_1+\ntbm F_2$, $\mu$ is the mobility and $2Dt$ is the dispersion in any direction of motion: $\l x_i(t) x_i (t)\r=2Dt$ with $D$ the diffusion coefficient.
The aforementioned Gaussian scenario is thus exact in this model with the role of $\dot w$ and $\dot q$ played by $\dot w_1=\ntbm F_1 \cdot \ntbm x/t$ and $\dot w_2=\ntbm F_2 \cdot \ntbm x/t$. One obviously has $\l\dot w_1\r= \mu \ntbm F_1 \cdot \ntbm F $, and $\l\dot w_2\r= \mu  \ntbm F_2 \cdot \ntbm F $. The corresponding correlation functions read $C_{11}=2D \|\ntbm F_1\|^2$, $C_{22}=2D \|\ntbm F_2\|^2$ and $C_{12}=2D \ntbm F_1 \cdot \ntbm F_2$. 
The large deviation of efficiency is given by  
\beq
J(\eta) = \frac{1}{2} \frac{(\eta \l\dot w_2\r+\l\dot w_1\r)^2}{\eta^2 C_{22} +2\eta C_{12}+ C_{11}}=\frac{\mu^2[(\eta \ntbm F_2+\ntbm F_1)\cdot \ntbm F]^2}{4D (\eta \ntbm F_2+\ntbm F_1)^2}.
\eeq
One immediately verifies that $J(\eta)$ takes its maximum value $J(1)=\mu^2 ||\bm{F}||^2/4D$ in $\eta=1$ which is the predicted reversible efficiency for work-to-work conversion. Furthermore, the above mentioned averages and correlations functions obey the relations (\ref{eq:ac}) upon setting $\eta_\nt{C}=1$ and $D=\mu T$ from the Einstein relation. One can thus also rewrite the $J(\eta)$ as in (\ref{eq:JetaCloseEq}) or (\ref{eq:scaling}) (with $\eta_\nt{C}=1$).

\subsection{Photo-electric device}
\begin{figure}
\vspace{0.4cm}
\includegraphics[width=6cm]{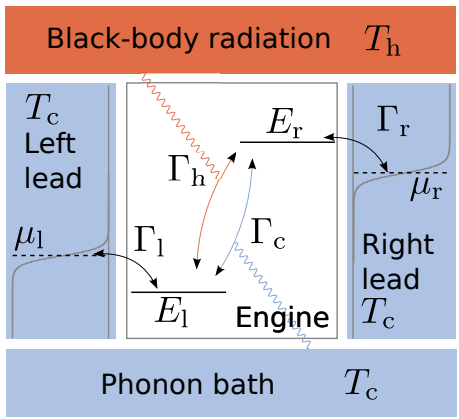}
\caption{Sketch of a photo-electric device. The device consists of two single level quantum dots (in white) connected to two leads (in blue) at temperature $T_\nt{c}$ and at different chemical potentials $\mu_\nt{l}$  and $\mu_\nt{r}$. The electron transitions between left and right quantum dots are induced either by photons from the black-body radiation at temperature $T_\nt{h}$ (in red) or by phonons at temperature $T_\nt{c}$ (in blue). The arrows indicate possible electronic transitions between different energy levels and the $\Gamma$'s represent the coupling strengths with the reservoirs.  \label{fig0}}
\end{figure}
\begin{figure}
\includegraphics[width=8.5cm]{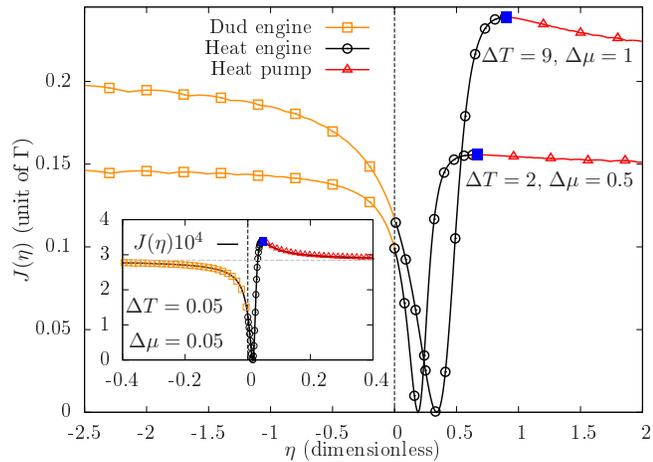}
\caption{Large deviation functions of efficiency $J(\eta)$. The curves are obtained from equation (\ref{eq:defJeta}) for the photo-electric device of Fig.~\ref{fig0} operating on average as
an heat engine. Each curve corresponds to a given temperature and chemical potential differences, black circles denote heat engine realizations, red triangles (resp. gold squares) denote heat pump (resp. a dud engine) realizations. The blue filled squares denote the least likely Carnot efficiency while the zeros correspondÒ to the most likely one. The left and right horizontal asymptotes coincide and correspond to realizations with low heat exchange. Inset: The close-to-equilibrium approximation (symbols) fits very well with the exact result (\ref{eq:JetaCloseEq}) (black solid line).
Parameters for the curves are $E_\nt{l}=0.5$, $E_\nt{r}=2.5$, $\mu_\nt{l}=1$, $T_\nt{c}=1$, $\Delta T = T_\nt{h} -T_\nt{c}$, $\Gamma_\nt{h}= \Gamma_\nt{l}=\Gamma_\nt{r} =10$ and $\Gamma_\nt{c} = 1$. \label{fig7}}
\end{figure}
Our second model is a nano-sized photoelectric device powered by black-body radiation at temperature $T_\nt{h}$ \cite{Rutten2009_vol80}. The device is composed of two quantum dots, each with a single energy level $E_\nt{l}$ and $E_\nt{r}$ ($E_\nt{r} > E_\nt{l}$), respectively, cf. Fig.~\ref{fig0}. Coulomb repulsion prevents simultaneous occupation by electrons of both quantum dots. Each dot can exchange electrons with its neighbouring electronic lead. Both leads are at the same temperature $T_\nt{c}$, but at different voltages and therefore at different chemical potentials $\mu_\nt{r} > \mu_\nt{l}$. Electron transfers between the two quantum dots are induced either by hot black-body radiation at $T_\nt{h}$ or by cold thermal phonons at $T_\nt{c}$. This device operates as an heat engine fuelled by the heat $q=n_p (E_\nt{r}-E_\nt{l})$, where $n_p$ is the number of photons absorbed from the hot black-body, and producing a positive work output $-w=n_e \Delta \mu$, where $n_e$ is the number of electrons transferred from left to right lead against the chemical potential gradient $\Delta \mu = \mu_\nt{r}-\mu_\nt{l} >0$. The stochastic efficiency is thus $\eta = -w/q$. The rates describing the Markovian dynamics of the device as well as the large deviation function for the work and heat statistics are discussed in the methods section. The resulting large deviation function for efficiency is plotted in Fig.~\ref{fig7}. All the predicted features - the least likely value at Carnot efficiency and the universal shape of the large deviation function close-to-equilibrium - are perfectly reproduced.

\section{Discussion}

The efficiency of macroscopic thermal machines is the ratio between two averaged quantities, the extracted work and the heat coming from the hot source. One of the momentous discoveries in science, which lead to the formulation of the second law of thermodynamics, is the observation by Carnot that this efficiency has a maximum called Carnot efficiency. Contrary to macroscopic machines, the behavior of small machines is subjected to strong fluctuations. Their average behavior thus provides an incomplete description except in the macroscopic limit where fluctuations are typically strongly peaked around the average. In the present letter, we introduce the concept of fluctuating efficiency to accurately characterize the performance of small machines and find universal features in its fluctuations. Using the fluctuation theorem, which generalizes the second law at the fluctuating level, we provide an analogue of the Carnot analysis by proving that the Carnot efficiency becomes the least likely efficiency when long measurement times are considered, independently of any details of the machine or of its mode of operation. Furthermore, we show that close-to-equilibrium the large deviation function of the efficiency fluctuations obeys a universal form parametrized by the Onsager matrix of the engine. Our study suggests a new direct application of the fluctuation theorem which was previously mostly invoked to measure free energy differences \cite{Jarzynski1997_vol78,Crooks2000_vol61,Alemany2012_vol8}. Since heat and work fluctuations are nowadays measured in a wide variety of systems \cite{Saira2012_vol109, Blickle2012_vol8, Moffitt2009_vol457, Yasuda2001_vol410, Toyabe2010_vol6, Berut2012_vol483, Alemany2012_vol8, Koski2013_vol9, Kung2012_vol2, Ciliberto2013_vol110, Bustamante2005_vol58, Matthews2013_vol}, we expect that experimental measurements of the fluctuating efficiency will become a valuable tool to characterize the performance of small engines.


\section{Methods}

\subsection{Linear response and fluctuation theorem} 

For the photoelectric device, the average photon and electron currents, $\dot N_\nt{p} \equiv \langle n_\nt{p} \rangle /t $ and $\dot N_\nt{e} \equiv \langle n_\nt{e} \rangle /t $, read in the linear regime 
\bea
\dot N_\nt{e} &=& L_\nt{ee}\Delta \mu /T_\nt{c} + L_\nt{ep} \Delta E \Delta \beta  , \label{eq:current} \\
\dot N_\nt{p} &=& L_\nt{ep}\Delta \mu /T_\nt{c} + L_\nt{pp} \Delta E \Delta \beta  ,
\label{eq:linearresponse}
\eea
where $\Delta \beta = 1/T_\nt{c} - 1/T_\nt{h} > 0$, $\Delta E = E_r - E_l > 0$ and $L$ is the symmetric Onsager matrix with $L_\nt{pp} \ge 0$, $L_\nt{ee} \ge 0$ and $\det L \ge 0$. The average work and heat per unit time can thus be written as
\bea
\dot W &=& \Delta \mu \dot N_\nt{e} =\frac{1 }{T_\nt{c}} \left( L_\nt{ee} \Delta \mu^2 + \eta_\nt{C} L_\nt{ep}\Delta \mu \Delta E \right), \label{Wdot}\\
\dot Q &=& \Delta E   \dot N_\nt{p} = \frac{1 }{T_\nt{c}} \left(  L_\nt{ep}\Delta \mu \Delta E + \eta_\nt{C} L_\nt{pp} \Delta E^2  \right). \label{Qdot}
\eea
From Green-Kubo relation, the linear response coefficients are related to equilibrium fluctuations by 
\beq
L_{ep}= \lim_{t \rightarrow \infty }\frac{1}{2t} \l [n_\nt{e}(t)-\langle n_\nt{e} \rangle_\nt{eq} ] [ n_\nt{p}(0) - \langle n_\nt{p} \rangle_\nt{eq}] \r_\nt{eq}.
\eeq
This implies that in the long time limit $ C_{wq}/2 \longrightarrow \Delta \mu \Delta E L_\nt{ep} $. Proceeding similarly for the other response coefficients we find $ C_{ww} /2\longrightarrow  \Delta \mu^2 L_\nt{ee} $ and $ C_{qq}/2 \longrightarrow \Delta E^2 L_\nt{pp}  $. Equations (\ref{Wdot}-\ref{Qdot}) thus lead to equation (\ref{eq:ac}) of the results section.
These equalities may also be derived using the fluctuation theorem for work and heat in the Gaussian limit. Indeed, using
\beq
I(\dot w, \dot q) - I(-\dot w,- \dot q) = -(\eta_\nt{C} \dot q + \dot w)\frac{1}{T_\nt{c}},
\label{eq:FTforLDFwq}
\eeq
and the quadratic large deviation function
\beq
I(\dot w, \dot q) = \frac{\left (\begin{array}{c}
\dot w-\dot W \\ \dot q -\dot Q 
\end{array} \right )^T
\left [\begin{array}{cc}
C_{qq} & -C_{wq}  \\
-C_{wq} & C_{ww}
\end{array} \right ]
\left (\begin{array}{c}
\dot w-\dot W \\
 \dot q -\dot Q 
\end{array} \right )}{2 \det C }, \label{eq:Iwq}
\eeq
we get
\beq
\frac{\det C}{2 T_\nt{c}} (\eta_\nt{C} \dot q+ \dot w) =  \dot w  \dot W  C_{qq} + \dot q   \dot Q C_{ww} - C_{wq} ( \dot w \dot Q + \dot q \dot W).
\eeq
Since this relation must hold true for any values of $\dot w $ and $ \dot q$, we obtain 
\beq
\frac{1}{\det C}\left [\begin{array}{cc}
C_{qq} & -C_{wq}  \\
-C_{wq} & C_{ww}
\end{array} \right ]\left (\begin{array}{c}
\dot W \\
\dot Q 
\end{array} \right ) = \left (\begin{array}{c}
1/2T_\nt{c}\\
\eta_\nt{C}/(2T_\nt{c})
\end{array} \right )
\eeq
which reproduces the expected result when solved for $\dot W$ and $\dot Q$.


\subsection{Photo-electric device: Heat and work statistics}

The work $w$ and heat $q$ statistics in the photo-electric device is obtained by considering the generating function 
$g_t(j,\gamma,\lambda) = \l e^{\gamma w + \lambda q} \r_j$ where the subscript $j$ denotes that the trajectory average is conditioned on the final state $j$ of the device at time $t$.
The three different states of the device are denoted $j=0,l,r$ for respectively no electrons in the device, one electron in the energy level $E_l$ connected to the left lead, or one electron in the energy level $E_r$ connected to the right lead. The generating function evolves according to
\begin{widetext}
\beq
\left (\begin{array}{c}
\dot g_t(0,\gamma,\lambda) \\
\dot g_t(l,\gamma,\lambda) \\
\dot g_t(r,\gamma,\lambda) 
\end{array} \right ) = \left [\begin{array}{ccc}
-k_{l0}-k_{r0} & k_{0l} & k_{0r} e^{-\gamma \Delta \mu} \\
k_{l0} & -k_{0l}-k_{rl} & k^{\nt c}_{lr}+k^{\nt h}_{lr}e^{-\lambda (E_r-E_l)} \\
k_{r0} e^{\gamma \Delta \mu} & k^{\nt c}_{rl}+k^{\nt h}_{rl}e^{\lambda (E_r-E_l)}  & - k_{0r} - k_{lr}
\end{array} \right ]  \left (\begin{array}{c}
g_t(0,\gamma,\lambda) \\
g_t(l,\gamma,\lambda) \\
g_t(r,\gamma,\lambda) 
\end{array} \right ) \label{eq:GenFuncEvo}.
\eeq
\end{widetext}
When $\lambda=\gamma=0$, (\ref{eq:GenFuncEvo}) becomes a Markovian master equation for the probability $P_j=g_t(j,0,0)$ to find the device in state $j$ at time $t$.
The rates $k_{ij}$ denote the probability per unit time to jump from state $j$ to $i$. 
Introducing the Fermi-Dirac distribution $f(x) \equiv 1/(e^{x}+1)$ and the Bose-Einstein distribution $b(x) \equiv 1/(e^{x}-1)$, they are defined by
\begin{align}
k_{l0} &\equiv \Gamma_l f \left ( \frac{E_l-\mu_l }{T_\nt{c}} \right ), &
k_{0l} &\equiv \Gamma_l \left [1-f\left ( \frac{E_l-\mu_l}{T_\nt{c}} \right ) \right ],\nonumber \\
k_{r0} &\equiv \Gamma_r f\left ( \frac{E_r-\mu_r}{T_\nt{c}}\right ), &
k_{0r} &\equiv \Gamma_r \left [1-f\left ( \frac{E_r-\mu_r}{T_\nt{c}}\right )\right ], \nonumber  \\
k^\nu_{rl} &\equiv \Gamma_\nu b\left (\frac{E_r-E_l}{T_\nu}\right ), &
k^\nu_{lr} &\equiv \Gamma_\nu \left [ 1 + b\left (\frac{E_r-E_l}{T_\nu}\right ) \right ],
\end{align}
and $k_{ij} \equiv k_{ij}^\nt{c}+ k_{ij}^\nt{h}$, where $\nu=\nt{c},\nt{h}$ denotes the cold and hot reservoir and the $\Gamma$'s the coupling strength with the various reservoirs \cite{Rutten2009_vol80} as illustrated in Fig.~\ref{fig0} of the result section. For long times $t$, the work and heat generating function is dominated by the highest eigenvalue $\phi(\gamma,\lambda)$ of the rate matrix in (\ref{eq:GenFuncEvo})
\beq
\l e^{\gamma w + \lambda q} \r = \sum_{j=0,l,r} g_t(j,\gamma,\lambda) \underset{t \ra \infty}{\sim} e^{t \phi(\gamma,\lambda)}.
\eeq
The latter can be calculated analytically. The corresponding large deviation function is obtained by the Legendre transform $I(\dot w, \dot q) = \max_{\gamma,\lambda} \{ \gamma \dot w  + \lambda \dot q - \phi(\gamma,\lambda)\}$. The large deviation function for efficiency fluctuations is obtained from it using equation (\ref{eq:defJeta}). Alternatively it can be obtained using $J(\eta) = -\min_\gamma \phi(\gamma,\eta \gamma)$. The proof will be provided in a forthcoming publication. This latter minimization has been performed numerically to produce Fig.~\ref{fig7} in the letter.

\subsection{Alternative expression of $J(\eta)$}

The three equations in the letter for $J(\infty)$, $J(\eta_\nt{C})$ and $\bar \eta$ expressed in term of the covariance matrix close to equilibrium can be inverted to obtain 
\begin{align}
C_{qq} &= \frac{ 8J(\eta_\nt{C})^2T_\nt{c}^2}{(\bar{\eta}-\eta_\nt{C})^2J(\infty)},  \\
C_{wq} &= -8J(\eta_\nt{C})T_\nt{c}^2 \frac{J(\infty)\bar{\eta}-J(\infty)\eta_\nt{C}+\eta_\nt{C}J(\eta_\nt{C})}{J(\infty)(\bar{\eta}^2-2\bar{\eta}\eta_\nt{C}+\eta_\nt{C}^2)},  \\
C_{ww} &= 8J(\eta_\nt{C})T_\nt{c}^2 \frac{J(\infty)\bar{\eta}^2+\eta_\nt{C}^2J(\eta_\nt{C})-J(\infty)\eta_\nt{C}^2}{J(\infty)(\bar{\eta}^2-2\bar{\eta}\eta_\nt{C}+\eta_\nt{C}^2)}. 
\end{align}
Using these coefficients, we recover equation (\ref{eq:scaling}) of the letter.

\section*{Acknowledgement}

G.V. acknowledges insightful comments from Andreas Engel.
This work was supported by the National Research Fund, Luxembourg under Project No. FNR/A11/02 and INTER/FWO/13/09 and also benefited from support by the ESF network ``Exploring the Physics of Small Devices''. 

\section*{Author contributions}

G.V. explicitly derived the central results of the letter and suggested to study efficiency fluctuations using large deviation theory. 
T.W. was involved in preliminary studies of model systems which lead to this work. 
C.V.d.B. supervised the work at every stage and provided key contributions about the connection between the fluctuation theorem and efficiency fluctuations.
M.E. supervised the work at every stage, proposed to study efficiency fluctuations and made key suggestions about the close-to-equilibrium limit. 

\blfootnote{$^1$ gatien.verley@gmail.com }

\bibliographystyle{nature}

\end{document}